\documentclass[sigconf]{acmart}


\usepackage{xcolor}
\usepackage[english]{babel}
\usepackage[utf8]{inputenc}
\usepackage{graphicx}
\usepackage{booktabs}
\usepackage{multicol}
\usepackage{multirow}
\usepackage{balance}
\usepackage{enumitem}

\usepackage{tcolorbox}

\usepackage[caption=false]{subfig}

\newcommand{\sbar}[1]{{\color{darkgray}\rule{\dimexpr 0.6cm * #1 / 100}{5pt}\color{lightgray}\rule{\dimexpr 0.6cm * (100 - #1) / 100}{5pt}}}

\newcommand{\datasetsize}[0]{6,785}
\newcommand{\totalDeprecatedProjects}[0]{248}
\newcommand{\totalActiveProjects}[0]{754}
\newcommand{\trainingsetsize}[0]{1,002}
\newcommand{\testsetsize}[0]{5,783}
\newcommand{\testsetClassifiedAsUnmaintained}[0]{2,856}

\newcommand{\testsetClassifiedAsActive}[0]{2,927}

\newcommand{\totalProjectsUnmaintainedByReadme}[0]{112}

\newcommand{\totalMaintainedProjectsBySurveyAnswers}[0]{20}
\newcommand{\totalFinishedProjectsBySurveyAnswers}[0]{54}
\newcommand{\totalDeprecatedProjectsBySurveyAnswers}[0]{41}
\newcommand{\totalOthersAnswersBySurvey}[0]{18}
\newcommand{\totalTruePositivesAnswersBySurvey}[0]{103}
\newcommand{\totalFalsePositivesAnswersBySurvey}[0]{26}
\newcommand{\totalProjectsWithCommitsInTheLastYear}[0]{77}

\newcommand{\totalSurveyParticipantsWithPublicEmail}{323}
\newcommand{\totalProjectsWithDeprecatedMessageInReadmeOfSurvey}{21}
\newcommand{\totalSurveyAnswersAndReadmeWithUnclearAnswers}[0]{133}
\newcommand{\totalSurveyAnswersAndReadme}[0]{129}
\newcommand{\totalSurveyAnswersByDevelopers}[0]{112}
\newcommand{\modelPrecision}[0]{80}
\newcommand{\modelRecall}[0]{96}

\usepackage{soul}
\usepackage{balance}
\usepackage{changepage}
\usepackage{framed}
\makeatletter
 {\par\unskip\endMakeFramed}
\makeatother

\definecolor{formalshade}{rgb}{0.93,0.93,0.93}

\definecolor{darkblue}{rgb}{0.2, 0.2, 0.2}

\newenvironment{formal}{%
  \def\FrameCommand{%
    \hspace{1pt}%
    {\color{darkblue}\vrule width 2pt}%
    {\color{formalshade}\vrule width 4pt}%
    \colorbox{formalshade}%
  }%
  \MakeFramed{\advance\hsize-\width\FrameRestore}%
  \noindent\hspace{-1pt}% disable indenting first paragraph
  \begin{adjustwidth}{}{7pt}%
  \vspace{2pt}\vspace{2pt}%
}
{%
  \vspace{3pt}\end{adjustwidth}\endMakeFramed%
}

\begin{document}

\settopmatter{printfolios=true}

\title{Identifying Unmaintained Projects in GitHub}
    
\author{Jailton Coelho}
\affiliation{Federal University of Minas Gerais, Brazil}
\email{jailtoncoelho@dcc.ufmg.br}

\author{Marco Tulio Valente}
\affiliation{Federal University of Minas Gerais, Brazil}
\email{mtov@dcc.ufmg.br}

\author{Luciana L.~Silva}
\affiliation{Federal Institute of Minas Gerais, Brazil}
\email{luciana.lourdes.silva@ifmg.edu.br}

\author{Emad Shihab}
\affiliation{Concordia University,Canada}
\email{eshihab@encs.concordia.ca}

\begin{abstract}

{\em Background:} Open source software has an increasing importance in modern software development. However, there is also a growing concern on the sustainability of such projects, which are usually managed by a small number of developers, frequently working as volunteers.
{\em Aims:} In this paper, we propose an approach to identify GitHub projects that are not actively maintained. Our goal is to alert users about the risks of using these projects and possibly motivate other developers to assume the maintenance of the projects.
{\em Method:} We train machine learning models to identify unmaintained or sparsely maintained projects, based on a set of features about project activity (commits, forks, issues, etc). We empirically validate the model with the best performance with the principal developers of \totalSurveyAnswersAndReadme\ GitHub projects.
{\em Results:} The proposed machine learning approach has a precision of \modelPrecision\%, based on the feedback of real open source developers; and a recall of \modelRecall\%. We also show that our approach can be used to assess the risks of projects becoming unmaintained.
{\em Conclusions:} The model proposed in this paper can be used by open source users and developers to identify GitHub projects that are not actively maintained anymore.

\end{abstract} 


\copyrightyear{2018} 
\acmYear{2018} 
\setcopyright{acmcopyright}
\acmConference[ESEM '18]{ACM / IEEE International Symposium on Empirical Software Engineering and Measurement (ESEM)}{October 11--12, 2018}{Oulu, Finland}
\acmBooktitle{ACM / IEEE International Symposium on Empirical Software Engineering and Measurement (ESEM) (ESEM '18), October 11--12, 2018, Oulu, Finland}
\acmPrice{15.00}
\acmDOI{10.1145/3239235.3240501}
\acmISBN{978-1-4503-5823-1/18/10}

\maketitle

\section{Introduction}
\label{sec:introduction}

Open source projects have an increasing relevance in modern software development~\cite{nadia2016roads}. For example, many critical software systems are currently available under open source licenses, including operating systems, compilers, databases, and web servers. Similarly, it is common nowadays to depend on open source libraries and frameworks when building and evolving proprietary software. For example, in a recent survey---conducted by Black Duck Software---78\% of the over 1,300 companies surveyed acknowledge the use of open source in their daily development.\footnote{\url{https://www.slideshare.net/blackducksoftware/2015-future-of-open-source-survey-results}} Concretely, Instagram---the popular photo-sharing social network---is currently implemented using more than 20 open source libraries.\footnote{\url{https://www.instagram.com/about/legal/libraries/}} Furthermore, the emergence of world-wide code sharing platforms---GitHub is the most well-known example---is contributing to transform open source development in a competitive market. Indeed, in a recent survey with open source maintainers we found that the most common reason for the failure of open source projects is the appearance of a stronger competitor in GitHub~\cite{coelho2017why}.

However, GitHub does not include clear data about project status, in terms of maintenance activity. Users can access historical data about commits or global project metrics, like number of stars, forks, and watchers. However, based on the values of theses metrics, they should judge themselves whether a project is being actively maintained (and therefore if it is worth to use it or not). Therefore, in this paper we propose and evaluate a machine learning approach to identify unmaintained (or sparsely maintained) projects in GitHub. Our goal is to provide a simple and effective mechanism to alert users about the risks of depending on a GitHub project. This information  can also contribute to attract new maintainers to a project. For example, users of libraries facing the risks of discontinuation can be motivated to assume their maintenance.

Previous work in this area relies on the last commit activity to classify projects as unmaintained or in similar status. For example, Khondhu et al. use an one-year inactivity threshold to classify {\em dormant} projects on SourceForge~\cite{khondhu2013all}. The same threshold is used in works by \citet{mens2014survivability}, \citet{izquierdo2017empirical}, and in our previous work about the motivations for the failure of open source projects~\cite{coelho2017why}. However, in this paper, we do not use this definition when investigating unmaintained projects due to three reasons. First, because defining a threshold to characterize unmaintained projects is not trivial. For example, in the mentioned works, this decision is arbitrary and it is not empirically validated. Second, our intention is to detect unmaintained projects as soon as possible; preferably, without having to wait for one year of inactivity. Third, our definition of unmaintained projects does not assume a complete absence of commits during a given period; instead, a project is considered unmaintained even when sporadic and few commits happen in a given time interval. Stated otherwise, by our definition, unmaintained projects do not necessarily need to be dead, deprecated or archived.

In this paper, we first train ten machine learning models to identify unmaintained projects, using  as features standard metrics provided by GitHub about a project's maintenance activity, e.g.,~number of commits, forks, issues, and pull requests. Then, we select the model with the best performance and validate it by means of a survey with the owners of projects classified as {\em unmaintained} and also with a set of deprecated GitHub projects. Particularly, we ask three research questions about properties of this model:\\[-.3cm]

\noindent{\em RQ1: What is the precision according to GitHub developers?} The intention is to check precision in the field, according to the feedback provided by the principal developers of popular GitHub projects.\\[-.3cm]

\noindent{\em RQ2: What is the recall when identifying unmaintained projects?} Recall is more difficult to compute in the field, because it requires the identification of all unmaintained projects in GitHub. To circumvent this problem, we compute recall considering only projects that declare in their README\footnote{READMEs are the first file a visitor is presented to when visiting a GitHub repository.} they are not under maintenance.\\[-.3cm]

\noindent{\em RQ3: How early does the  model identify unmaintained projects?} As mentioned, the proposed model does not depend on an inactivity interval to classify a project as unmaintained. Therefore, in this  question, we investigate whether this ability is  effective in the field, when identifying the maintenance status of real GitHub projects.

Our contributions are twofold: (1) we propose a machine learning approach to identify unmaintained (or sporadically maintained) projects in GitHub, which achieved a precision of \modelPrecision \% and a recall of \modelRecall \% when validated with real open source developers and projects; (2) we propose a metric to reveal the maintenance activity level of GitHub projects.

This paper is organized as follows. In Section~\ref{sec:machine-learning}, we present and evaluate a machine learning model to identify unmaintained projects. Section~\ref{sec:validation} validates this model with GitHub developers and  projects that are documented as deprecated. Section~\ref{sec:level-of-maintenance-activity} defines and discusses the Level of Maintenance Activity (LMA) metric. Section~\ref{sec:threats} lists threats to validity and Section~\ref{sec:related-work} discusses related work. Section~\ref{sec:conclusion} concludes the paper and outlines further work.

\section{Machine Learning Model}
\label{sec:machine-learning}

In this section, we describe our machine learning approach to identify projects that are no longer under maintenance.

\subsection{Experimental Design}
\label{sec:prediction-study-design}

\begin{table*}[!ht]
    \centering
    \caption{Features used to identify unmaintained projects.}   
    \begin{tabular}{ l l l}
        \toprule
        {\bf Dimension} 				& {\bf Feature}			& {\bf Description}			\\ 
        \midrule
        \multirow{9}{*}{\bf Project} 
        &	Forks							& 	Number of forks created by developers	\\
        &	Open issues						& 	Number of issues opened by developers	\\     
        &	Closed issues					& 	Number of issues closed by developers	\\
        &	Open pull requests 				& 	Number of pull requests opened by the project developers\\
        &	Closed pull requests			& 	Number of pull requests closed by the project developers\\
        &	Merged pull requests			& 	Number of pull requests merged by the project developers\\
        &	Commits							& 	Number of commits performed by developers		\\
        & 	Max days without commits		& 	Maximum number of consecutive days without commits	\\
        & 	Max contributions by developer	& 	Number of commits of the developer with the highest number of commits\\
        \midrule
        \multirow{2}{*}{\bf  Contributor} 
        &	New contributors				& 	Number of contributors who made their first commit in the considered period	\\
        & 	Distinct contributors			& 	Number of distinct contributors that committed in the considered period	\\
        \midrule
        \multirow{2}{*}{\bf  Owner} 
        &	Projects created by the owner	& 	Number of projects created by a given owner \\
        &	Number of commits of the owner	& 	Number of commits performed by a given owner  	\\      
        \bottomrule
    \end{tabular}
    \label{tab:feature-means}
\end{table*}

\noindent{\bf Dataset.} We start with a dataset containing the top-10,000 most starred projects on GitHub (in November, 2017). Stars---GitHub's equivalent for {\em likes} in other social networks---is a common proxy for the popularity of GitHub projects~\cite{borges2016icsme}. Then, we follow three strategies in order, to discard projects from this initial selection. First, we remove 2,810 repositories that have less than two years from the first to the last commit (because we need historical data to compute the features used by the prediction models). Second, we remove 331 projects with null size, measured in lines of code (typically, these projects are implemented in non-programming languages, like CSS, HTML, etc). Finally, we remove 74 non-software projects, which are identified by searching for the following topics: {\em books} and {\em awesome-lists}.\footnote{GitHub topics allow tagging a repository with keywords.} We end up with a list of \datasetsize\ projects. 

Next, we define two subsets of systems: {\em active} and {\em unmaintained}. The active (or under maintenance) group is composed by \totalActiveProjects\ projects that have at least one release in the last month, including well-known projects, like {\sc facebook/react}, {\sc d3/d3}, and {\sc nodejs/node}. Thus, we assume that projects with recent releases are active (under maintenance). By contrast, the unmaintained group is composed by \totalDeprecatedProjects\ projects, including 104 projects that were explicitly declared by their principal developers as unmaintained in our previous work~\citep{coelho2017why} and 144 {\em archived} projects. Archiving is a recent feature provided by GitHub that allows developers to explicitly move their projects to a read-only state. In this state, users cannot create issues, pull requests, or comments, but can still fork or star the projects.\\[-.2cm]

\begin{figure}[!ht]
\centering
\includegraphics[width=8.5cm]{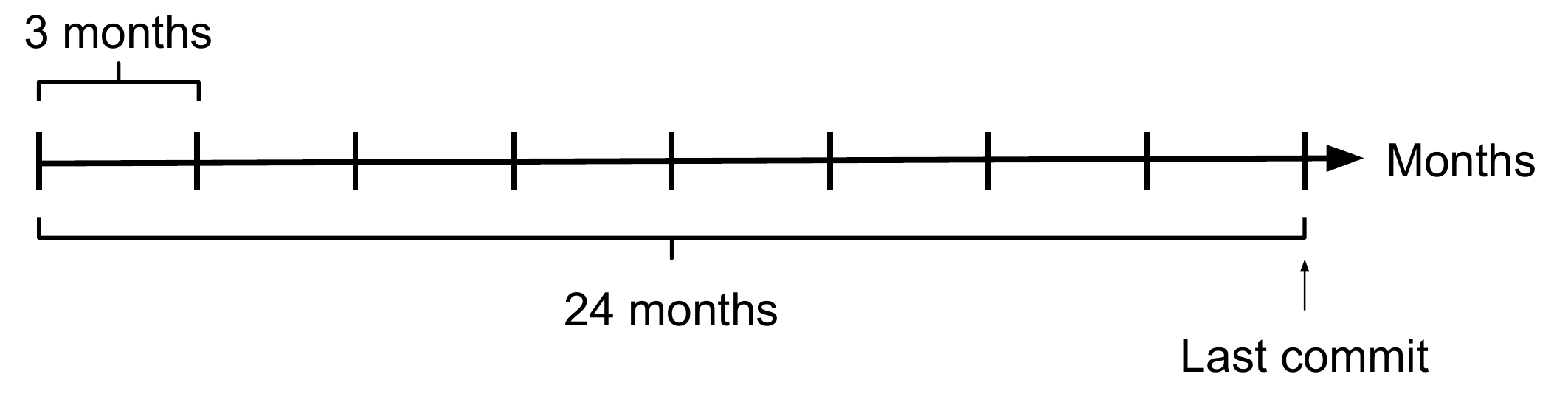}
\caption{Feature collection during 24 months in 3-month intervals.}
\label{fig:intervals}
\end{figure}

\noindent{\bf Features}. Our hypothesis is that a machine learning classifier can identify unmaintained projects by considering features about (1) projects, including number of forks, issues, pull requests, and commits; (2) contributors, including number of new and distinct contributors (the rationale is that maintenance activity might increase by attracting new developers); (3) project owners, including number of projects he/she owns and total number of commits in GitHub (the rationale is that maintenance might be affected when project owners have many projects on GitHub). In total, we consider 13 features, as described in Table~\ref{tab:feature-means}.
The feature values are collected using GitHub's official API. However, they do not refer to the whole history of a project, but only to the last $n$ months, counting from the last commit; moreover, we collect each feature in intervals of $m$ months. The goal is to derive temporal series of feature values, which can be used by a machine learning algorithm to infer trends in the project evolution, e.g., an increasing number of opened issues or a decreasing number of commits. Figure~\ref{fig:intervals} illustrates the feature collection process assuming that $n$ is 24 months and that $m$ is 3 months. In this case, for each feature, we collect 8 data points, i.e.,~feature values.

We experiment with different combinations of $n$ and $m$; each one is called a scenario, in this paper. Table~\ref{tab:scenarios} describes the total number of data points extracted for each scenario. This number ranges from 13 data points (scenario with features extracted in a single interval of 6 months) to 104 data points  (scenario with features extracted in intervals of 3 months during 24 months, as in Figure~\ref{fig:intervals}).

\begin{table}[!h]
    \centering
    \caption{Scenarios used to collect features and train the machine learning models (length and intervals are in months; data points is the total number of data points collected for each scenario).}  
 	\small
    \begin{tabular}{l r r r r r r r r r r}
     \toprule
	\textbf{Scenario} & \textbf{1} & \textbf{2} & \textbf{3} & \textbf{4} & \textbf{5} & \textbf{6} & \textbf{7} & \textbf{8} & \textbf{9} & \textbf{10} \\
	\hline
    \textbf{Length} 	 & 6  & 6  & 12  & 12  & 12  & 18  & 18  & 24  & 24  & 24 \\
    \textbf{Intervals} 	 & 3  & 6  & 3   & 6   & 12  & 3   & 6   & 3   & 6   & 12 \\
    \textbf{Data points} & 26 & 13 & 52  & 26  & 13  & 78  & 39  & 104 & 52  & 26  \\

    \bottomrule
    \end{tabular}
    \label{tab:scenarios}
\end{table}

\begin{figure}[!t]
\centering
\includegraphics[width=8.5cm]{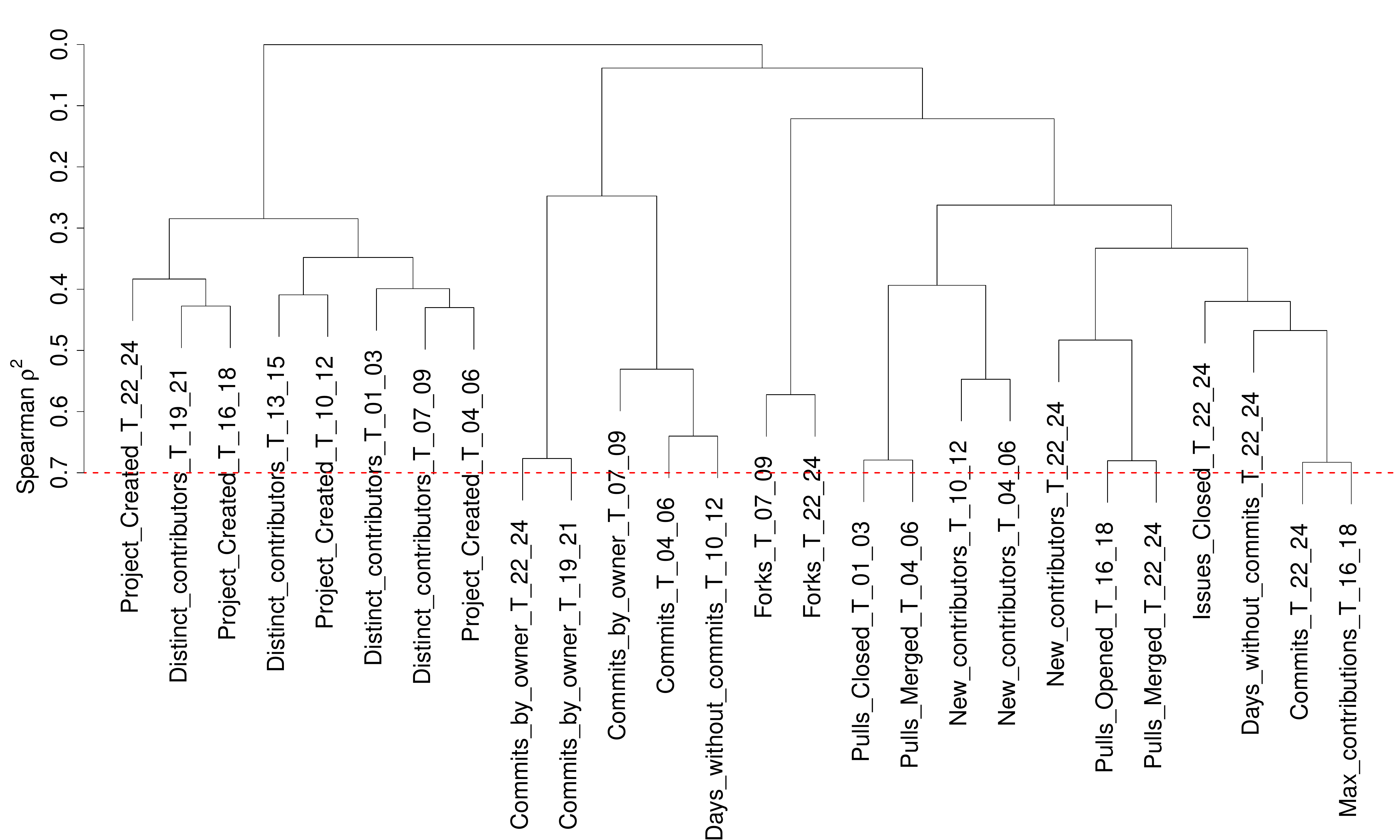}
\caption{Correlation analysis for the 104 data points collected for the features in scenario 8 (24 months, 3-month interval). 78 data points (75\%) are removed in this case, due to correlations with other data points, and therefore they do no appear in this final clustering.}
\label{fig:dendogram}
\end{figure}

\noindent{\bf Correlation Analysis.} As usual in machine learning experiments, we remove correlated features, following the process described by \citet{bao2017will}. To this purpose, we use a clustering analysis---as implemented in a R package named {\em Hmisc}\footnote{http://cran.r-project.org/web/packages/Hmisc/index.html}---to derive a hierarchical clustering of correlations among data points (extracted for the features in each scenario). For sub-hierarchies  with correlations larger than 0.7, we select only one data point for inclusion in the respective machine learning model, as common in other works~\citep{bao2017will, tian2015characteristics}.  
For example, Figure~\ref{fig:dendogram} shows the final hierarchical clustering for the scenario with 24 months, considering a 3-month interval (scenario~8). The  analysis in this scenario checks correlations among 104 data points (13 features $\times$ 8 data points per feature). As a result, 78 data points are removed due to correlations with other points and therefore do not appear in the dendogram presented in Figure~\ref{fig:dendogram}. Finally, Table~\ref{tab:features-removed} shows the total number and percentage of data points removed in each scenario, after correlation analysis. As we can see, the percentage of removed  points is relevant, ranging from 43\% (scenario 7) to 75\% (scenario 8).\\[-.2cm]

\begin{table}[!h]
    \centering
    \caption{Total number and percentage of data points removed in each scenario, after correlation analysis.}    
    \begin{tabular}{c r r r r r r r r r r}
     \toprule
	\textbf{Scenario} & \textbf{1} & \textbf{2} & \textbf{3} & \textbf{4} & \textbf{5} & \textbf{6} & \textbf{7} & \textbf{8} & \textbf{9} & \textbf{10} \\
	\hline
	%Period 	& 2  	&  2 & 2  &	1.5 & 1.5 & 1  & 1  & 1  & 0.5 & 0.5   \\ 
	%Interval 	& 3 	& 6  & 12 & 3   & 6   & 3  & 6  & 12 & 3   & 6   \\ 
	%Total 		& 104 	& 52 & 26 & 78  & 39  & 52 & 26 & 13 & 26  & 13   \\ 
    \# & 17 & 6  & 38 & 18 & 7  & 56 & 17 & 78 & 34 & 19   \\ 
    \% & 65 & 46 & 73 & 69 & 54 & 72 & 43 & 75 & 65 & 73   \\
    %\hline
        \bottomrule
    \end{tabular}
    \label{tab:features-removed}
\end{table}

\begin{table*}[!ht]
    \centering
    \caption{Prediction results (mean  of 100 iterations, using 5-cross validation); best results are in bold.}  
    \small
    \begin{tabular}{ l | r r | r r r | r r | r r r }
        \toprule
        \multirow{2}{*}{\bf Metrics} 
        & \multicolumn{2}{c | }{{\bf 0.5 Year}}
        & \multicolumn{3}{c | }{{\bf 1 Year}}
        & \multicolumn{2}{c | }{{\bf 1.5 Years}}
        & \multicolumn{3}{c  }{{\bf 2 Years}} \\
    	\multicolumn{1}{c | }{}
        & \multicolumn{1}{c|}{\textbf{3 months}}	
        & \multicolumn{1}{c|}{\bf 6 months}
        & \multicolumn{1}{c|}{\bf 3 months}
        & \multicolumn{1}{c|}{\bf 6 months}
        & \multicolumn{1}{c|}{\bf 12 months}         
        & \multicolumn{1}{c|}{\bf 3 months}
        & \multicolumn{1}{c|}{\bf 6 months} 
        & \multicolumn{1}{c|}{\bf 3 months}    
        & \multicolumn{1}{c|}{\bf 6 months}
        & \multicolumn{1}{c}{\bf 12 months} 
        \\
        \midrule	%		0.5/3		0.5/6	  1/3	 1/6	1/12  1.5/3	  1.5/6	  2/3	2/6  2/12
        Accuracy	&  \sbar{90} {0.90} & \sbar{91} {0.91} & \sbar{91} {0.91} & \sbar{90} {0.90} & \sbar{89} {0.89} & \sbar{91} {0.91} & \sbar{90} {0.90} & \sbar{92} {\bf 0.92} & \sbar{91} {0.91} & \sbar{90} {0.90} \\
        Precision	&  \sbar{83} {0.83} & \sbar{87} {0.87} & \sbar{87} {0.87} & \sbar{84} {0.84} & \sbar{82} {0.82} & \sbar{86} {0.86} & \sbar{83} {0.83} & \sbar{85} {\bf 0.86} & \sbar{85} {0.85} & \sbar{83} {0.83} \\
        Recall		&  \sbar{78} {0.78} & \sbar{74} {0.74} & \sbar{77} {0.77} & \sbar{75} {0.75} & \sbar{72} {0.72} & \sbar{78} {0.78} & \sbar{76} {0.76} & \sbar{81} {\bf 0.81} & \sbar{79} {0.79} & \sbar{73} {0.73} \\
        F-measure 	&  \sbar{80} {0.80} & \sbar{79} {0.79} & \sbar{81} {0.81} & \sbar{79} {0.79} & \sbar{77} {0.77} & \sbar{82} {0.82} & \sbar{79} {0.79} & \sbar{83} {\bf 0.83} & \sbar{82} {0.82} & \sbar{78} {0.78} \\
        Kappa		&  \sbar{74} {0.74} & \sbar{74} {0.74} & \sbar{76} {0.76} & \sbar{73} {0.73} & \sbar{70} {0.70} & \sbar{76} {0.76} & \sbar{73} {0.73} & \sbar{78} {\bf 0.78} & \sbar{76} {0.76} & \sbar{71} {0.71} \\
        AUC			&  \sbar{86} {0.86} & \sbar{85} {0.85} & \sbar{86} {0.86} & \sbar{85} {0.85} & \sbar{83} {0.83} & \sbar{87} {0.87} & \sbar{85} {0.85} & \sbar{88} {\bf 0.88} & \sbar{87} {0.87} & \sbar{84} {0.84} \\
        \bottomrule
    \end{tabular}
    \label{tab:prediction-results-scenarios}
\end{table*}

\noindent{\bf Machine Learning Classifier.} 
We use the data points extracted in each scenario to train and test models for predicting whether a project is unmaintained. In other words, we train and test ten machine learning models, one for each scenario. After that, we select the best model/scenario to continue with the paper. Particularly, we use the Random Forest algorithm~\citep{breiman2001random} to train the models because it has several advantages, such as robustness to noise and outliers~\citep{tian2015characteristics}. In addition, it is adopted in many other software engineering works~\citep{menzies2013local, peters2013better,  provost2001robust,fse2016-andre}. We compare the result of Random Forest with two baselines: baseline \#1 (all projects are predicted as unmaintained) and baseline \#2 (random predictions). We use the Random Forest implementation provided by {\em randomForest}'s R package\footnote{https://cran.r-project.org/web/packages/randomForest/} and 5-fold stratified cross validation to evaluate the models effectiveness. In 5-fold cross validation, we randomly divide the dataset into five folds, where four folds are used to train a classifier and the remaining fold is used to test its performance. Specifically, stratified cross validation is a variant, where each fold has approximately the same proportion of each class~\citep{breiman2001random}. We perform 100 rounds of experiments and report average results.\\[-.2cm]

\noindent{\bf{Evaluation Metrics.}}  
When evaluating the projects in the test fold, each project has four possible outcomes: (1) it is truly classified as unmaintained (True Positive); (2) it is classified as unmaintained but it is actually an active project (False Positive); (3) it is classified as an active project but it is actually an unmaintained one (False Negative); and (4) it is truly classified as an active project (True Negative). Considering these possible outcomes, we use six metrics to evaluate the performance of a classifier: precision, recall, F-measure, accuracy, AUC (Area Under Curve), and Kappa, which are commonly adopted in machine learning studies~\citep{tian2015characteristics, tian2015automated, da2014empirical, lamkanfi2010predicting, lessmann2008benchmarking}. Precision and recall measure the correctness and completeness of the classifier, respectively. F-measure is the harmonic mean of precision and recall. Accuracy measures how many projects are classified correctly over the total number of projects. AUC refers to the area under the Receiver Operating Characteristic (ROC) curve. Finally, kappa evaluates the relationship between the observed accuracy and the expected one~\cite{bookML_R}, which is particularly relevant in imbalanced datasets, as the dataset used to train the machine learning models (\totalActiveProjects\ active projects {\em vs} \totalDeprecatedProjects\ unmaintained ones).

\subsection{Experimental Results} 
Table~\ref{tab:prediction-results-scenarios} shows the results for each scenario. As we can see, Random Forest has the best results (in bold) when the features are collected during 2 years, in intervals of 3 months.  In this scenario, precision is 86\% and recall is 81\%, leading to a F-measure of 83\%. Kappa is 0.78---usually, kappa values greater than 0.60 are considered quite representative~\cite{kappa}. Finally, AUC is 0.88, which is an excellent result in the Software Engineering domain~\citep{lessmann2008benchmarking, thung2012automatic, tian2015characteristics}. Table~\ref{tab:comparison-baselines} compares the results of the best scenario/model with baseline~\#1 (all projects are predicted as unmaintained) and baseline~\#2 (random predictions). Despite the baseline under comparison, there are major differences in all evaluation metrics. For example, F-measure is 0.37 (baseline \#1) and 0.30 (baseline \#2), against 0.83 (proposed model).

\begin{table}[!ht]
    \centering
    \caption{Comparison of the proposed machine learning model with  baseline \#1 (all projects are predicted as unmaintained) and baseline \#2 (random predictions).}    
    \begin{tabular}{ l r r r r}
        \toprule
       {\bf Metrics}    & \multicolumn{1}{c}{\bf Model} &  \multicolumn{1}{c}{\bf Baseline \#1}  &  \multicolumn{1}{c}{\bf Baseline \#2}\\ 
        \midrule
        Accuracy	   	& 	\sbar{92} {0.92}  	& \sbar{22} {0.22} 	& \sbar{49} {0.49} \\
        Precision	   	&  	\sbar{86} {0.86}	& \sbar{22} {0.22} 	& \sbar{22} {0.22} \\
        Recall	       	&  	\sbar{81} {0.81}	& \sbar{100} {1.00} & \sbar{48} {0.48}	\\
        F-measure 	   	&  	\sbar{83} {0.83}	& \sbar{37} {0.37} 	& \sbar{30} {0.30} \\
        Kappa		   	&	\sbar{78} {0.78}	& \sbar{0} {0.00} 	& \sbar{1} {0.01} \\     
        AUC		   		&	\sbar{88} {0.88}	& \sbar{50} {0.50} 	& \sbar{49} {0.49} \\   
        \bottomrule
    \end{tabular}
    \label{tab:comparison-baselines}
\end{table}

Random Forest produces a measure of the importance of the predictor features. Table~\ref{tab:features-importance} shows the top-5 most important features by Mean Decrease Accuracy (MDA), for the best model. Essentially, MDA measures the increase in prediction error (or reduction in prediction accuracy) after randomly shifting the feature values~\cite{calle2010letter,louppe2013understanding}.
As we can see, the most important feature is the number of commits in the last time interval (i.e., the interval delimited by months 22-24, $T_{22,24}$), followed by the maximal number of days without commits in the same interval and in the interval $T_{10,2}$.
As also presented in Table~\ref{tab:features-importance}, the first four features are related to commits; the first feature non-related with commits is the number of issues closed in the first time interval ($T_{1,3}$).
\begin{table}[!h]
    \centering
    \caption{Top-5 most relevant features, by Mean Decrease Accuracy (MDA).}    
    \begin{tabular}{ l l r }
        \toprule
        {\bf Feature}					& {\bf Period}		& {\bf MDA} 	\\ 
        \midrule
        Commits							& T$_{22,24}$		& 	38.5	\\
        Max days without commits		& T$_{22,24}$		& 	28.6	\\
        Max days without commits		& T$_{10,12}$		& 	21.9	\\
        Max contributions by developer	& T$_{16,18}$		& 	21.1	\\
        Closed issues					& T$_{1,3}$			& 	18.0	\\
        \bottomrule
    \end{tabular}
    \label{tab:features-importance}
\end{table}

\section{Empirical Validation}
\label{sec:validation}

In this section, we {\em validate} the proposed machine learning model by means of a survey with the owners of projects classified as {\em unmaintained} and also with a set of deprecated GitHub projects.
Overall, our goal is to strengthen the confidence on the practical value of the model proposed in this work. Particularly, we provide answers to three research questions about this model:\\[-.3cm]

\noindent{\em RQ1: What is the precision according to GitHub developers?} \\[-.3cm]

\noindent{\em RQ2: What is the recall when identifying deprecated  projects?} \\[-.3cm]

\noindent{\em RQ3: How early does the  model identify unmaintained projects?} 

\subsection{Methodology}

\noindent{\bf RQ1:} To answer RQ1, we conduct a survey with GitHub developers. To select the participants, we first apply the proposed machine learning model in all projects from our dataset that were not used in the model's construction, totaling \testsetsize\ projects (\datasetsize\ $-$ \trainingsetsize\ projects). Then, we select \testsetClassifiedAsUnmaintained\ projects classified as unmaintained by the proposed model. From this sample, we remove 264 projects whose developers were recently contacted in our previous surveys~\cite{coelho2017why,coelho2018why}. We make this decision to not bother again these developers, with new e-mails and questions. Finally, we remove 2,270 projects whose owners do not have a public e-mail address on GitHub. As a result, we obtain a list of \totalSurveyParticipantsWithPublicEmail\ survey participants (2,856 $-$ 2,270 $-$ 264). However, before e-mailing these participants, the first author inspected the main page of each project on GitHub, to check whether it includes mentions to the project status, in terms of maintenance. We found \totalProjectsWithDeprecatedMessageInReadmeOfSurvey\ projects whose documentation states they are no longer maintained, by means of messages like this one:\\[-.3cm]

\noindent {\em This project is deprecated. It will not receive any future updates or bug fixes. If you are using it, please migrate to another solution.}\\[-.3cm]

Therefore, we do not send mails to the project owners, in such cases; and automatically consider these \totalProjectsWithDeprecatedMessageInReadmeOfSurvey\ projects as {\em unmaintained}.\\[-.2cm]

\noindent{\em Survey Period:} 
The survey was performed in the first two weeks of May, 2018. It is important to highlight that the machine learning model was constructed using data collected on November, 2017. Therefore, the {\em unmaintained} predictions evaluated in the survey refer to this date. We wait five months to ask the developers about the status of their projects because it usually takes some time until  developers actually accept the unmaintained condition of their projects. In other words, this section is based on predictions performed and available on November, 2017. However, these predictions are validated five months later, on May, 2018.\\[-.2cm]

\noindent{\em Survey Pilot and Questions:}  Initially, we perform a pilot survey with 75 projects ($\approx$ 25\%), randomly selected among the remaining 302 projects (\totalSurveyParticipantsWithPublicEmail\ $-$ \totalProjectsWithDeprecatedMessageInReadmeOfSurvey\ projects). We e-mail the principal developers of these projects with a single open question, asking them to confirm (or not) if their projects are unmaintained. We received 23 answers, which corresponds to a response ratio of 30.6\%. Then, two authors of this paper analyzed together the received answers to derive a list of recurrent themes. As a result, the following three common maintainability status were identified:\footnote{Project names are omitted, to preserve the respondent's anonymity; survey participants are identified by means of labels, in the format Pxx, where {\em xx} is an integer.}

\begin{enumerate}[leftmargin=*]

\item {\bf My project is under maintenance and new features are under implementation (6 answers):} As an example, we can mention the following answer:\\[-.3cm]

\noindent{\em [Project-Name] is still maintained. I maintain the infrastructure side of the project myself (e.g., make sure it's compatible with the latest Ruby version, coordinate PRs and issues, mailing list, etc.) while community provides features that are still missing. One such big feature is being developed as we speak and will be the highlight of the next release.} (P57)\\[-.3cm]

\item {\bf My project is finished and I only plan to fix important bugs (13 answers):} As an example, we mention the following answers:\\[-.3cm]

\noindent{\em It's just complete, at least for now. I still fix bugs on the rare occasion they are reported.} (P10)\\[-.3cm]

\noindent{\em I view it as basically ``done''. I don't think it needs any new features for the foreseeable future, and I don't see any changes as urgent unless someone discovers a major security vulnerability or something. I will probably continue to make changes to it a couple times per year, but mostly bugfixes.} (P68)\\[-.3cm]

\item {\bf My project is deprecated and I do not plan to implement new features or fix bugs (4 answers):} As an example, we can mention the following answer:\\[-.3cm]
 
\noindent{\em The project is unmaintained and I'll archive it.} (P74)

\end{enumerate}

After the pilot study, we proceed with the survey, by e-mailing the remaining 227 projects. However, instead of asking an open question---as in the pilot---we provide an objective question to the survey participants, about the maintainability status of their projects. In this objective question, we ask the participants to select one (out of three) status identified in the pilot study, plus an {\em other} option. This fourth option also includes a text form for the participants detailing their answers, if desired. Essentially, we change to an objective question format to make answering the survey easier, but without limiting the respondents' freedom to provide a different answer from the listed ones. In this final survey, we received 89 answers, representing a response ratio of 39.2\%. When considering both phases (pilot and final survey), we sent 302 e-mails, received \totalSurveyAnswersByDevelopers\ answers, representing an overall response ratio of 37.1\%. After adding the \totalProjectsWithDeprecatedMessageInReadmeOfSurvey\ projects that document their maintainability status, this empirical validation is based on \totalSurveyAnswersAndReadmeWithUnclearAnswers\ projects.\\[-.2cm]

\noindent{\bf RQ2:} To answer this second question, we construct a ground truth with projects that are no longer being maintained. First, we build a script to download the README (the main page of GitHub's projects) and search for a list of sentences that are commonly used to declare the unmaintained state of GitHub projects. This list is reused from our previous work~\cite{coelho2017why}, where we study the motivations for the failure of open source projects. It includes 32 sentences; in Table~\ref{tab:sentences}, we show a partial list, with 15 sentences.

\begin{table}[!h]
    \centering
    \caption{Sentences documenting unmaintained projects.}  
 
    \begin{tabular}{l}
     \toprule
{\em no longer under development}, 
{\em no longer supported or updated},\\
    
{\em deprecation notice},
{\em dead project},
{\em deprecated},
{\em unmaintained},
\\
{\em no longer being actively maintained},
{\em not maintained anymore},
\\
{\em not under active development},
{\em no longer supported},
\\
{\em is not supported},
{\em is not more supported},
{\em no longer supported},
\\
{\em no new features should be expected},
{\em isn’t maintained anymore}\\
\bottomrule
    \end{tabular}
    \label{tab:sentences}
\end{table}

We searched (in May, 2018) for these sentences in the README of \testsetsize\ projects, which represent all \datasetsize\ projects selected for this work minus \trainingsetsize\ projects used in Section~\ref{sec:machine-learning}. In 451 READMEs (7.8\%) we found the mentioned sentences. Then, the first author of this paper carefully inspected each README, to confirm the sentences indeed refer to the project's status, in terms of maintenance. In the case of \totalProjectsUnmaintainedByReadme\ projects ($\approx$ 25\%), he confirmed this fact. Therefore, these projects are considered as a ground truth for this investigation. Usually, the unconfirmed cases refer to the deprecation of specific elements, e.g., methods or classes.%\\[-.2cm]

\noindent{\bf RQ3:} To answer this third research question, we rely on projects whose unmaintained status, as predicted by the proposed model, is confirmed by the participants of the survey conducted in RQ1. Then, we compute  the number of days between November 30, 2018 (when the machine learning model proposed in this paper was built) and the last commit of the mentioned projects. For projects where this interval is less than one year, the proposed model is better than the strategy adopted in previous work~\cite{khondhu2013all,mens2014survivability, izquierdo2017empirical, coelho2017why}, which requires one year of commit inactivity to identify unmaintained projects.

\subsection{Results}

\noindent{\bf \large RQ1: Precision according to GitHub developers}\\[-.2cm]

\begin{figure}[!t]
\centering
\includegraphics[width=8cm]{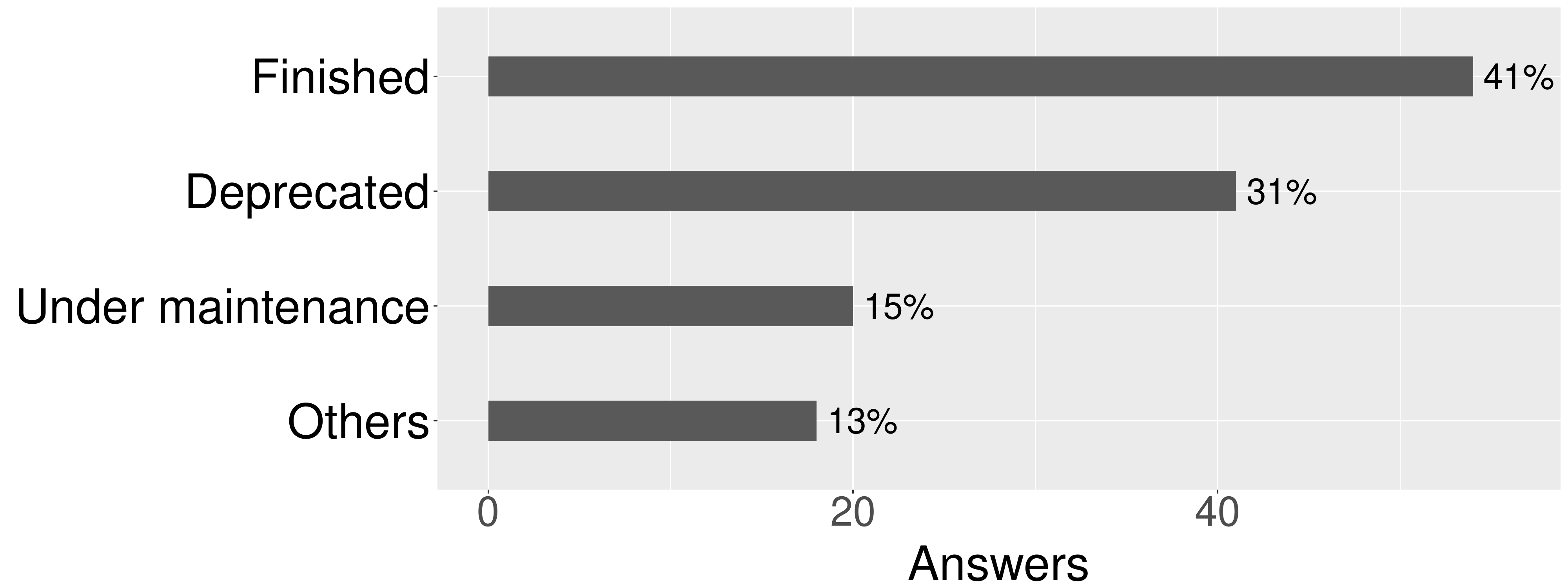}
\caption{Survey answers about projects' status.}
\label{fig:survey_answers}
\end{figure}

Before presenting the precision results, Figure~\ref{fig:survey_answers} shows the survey results, including answers retrieved from the project's documentation, answers received in the pilot and answers received in the final survey. As we can see, the most common status, with \totalFinishedProjectsBySurveyAnswers\ answers (41\%) refers to {\em finished} projects, i.e., cases where maintainers see their projects as feature-completed and only plan to resume the maintenance work if relevant bugs are reported.\footnote{In our previous work~\cite{coelho2017why}, we also identified finished or completed open source projects. However, we argued these projects do not contradict Lehman's Laws about software evolution~\cite{lehman1980programs}, because they usually deal with a very stable or controlled environment (whereas Lehman's Laws focus on E-type systems, where {\em E} stands for evolutionary).} We also received \totalDeprecatedProjectsBySurveyAnswers\ answers (31\%) mentioning the projects are deprecated and no further maintenance is planned, including the implementation of new features and bug fixes. Finally, we received \totalOthersAnswersBySurvey\ answers in the {\em other} option. In this case, four participants did not describe their answer or provide unclear answers; furthermore, one participant mentioned his project is in a {\em limbo} state:\\[-.3cm]

\noindent{\em The status of [Project-Name] fits into a special category. Some of the tools it’s based on are either deprecated or not powerful enough for the goal of the project. This is part of the reason what’s keeping the project from being ``done''. I would call this status {\bf \em limbo}. (P24)}\\[-.3cm]

Seven participants answered their projects are {\em stalled}, as in this answer:\\[-.3cm]

\noindent{\em It is under going a rewrite... but has been {\bf \em stalled} based on my own priorities (P33)}\\[-.3cm]

To compute precision, we consider as {\em true positive} answers related to the following status: finished (\totalFinishedProjectsBySurveyAnswers\ answers), deprecated (\totalDeprecatedProjectsBySurveyAnswers\ answers), stalled (7 answers), and limbo (1 answer). The remaining answers are interpreted as {\em false positives}, including answers mentioning that new features are being implemented (\totalMaintainedProjectsBySurveyAnswers\ answers) and the answers associated to the fourth option ({\em other} option), but without including a description or with an unclear description (4 answers). By following this criteria, we received \totalTruePositivesAnswersBySurvey\ true positive answers and \totalFalsePositivesAnswersBySurvey\ false positive ones, resulting in a precision of \modelPrecision\%.\footnote{This computation of precision assumes that finished projects are unmaintained. However, we recognize that the risks of using finished projects might be lower than the ones faced when using deprecated or stalled projects.}

\begin{formal}
By validating the proposed model with \totalSurveyAnswersAndReadme\ GitHub developers, we achieve a {\bf precision} of \modelPrecision\%, which is a result very close to the one obtained when building the model (86\%). 
\end{formal}

\vspace*{0.2cm}
\noindent{\bf \large RQ2: Recall considering deprecated projects}\\[-.2cm]

The proposed machine learning model classifies 108 (out of \totalProjectsUnmaintainedByReadme) projects of the constructed ground truth as unmaintained, which represents a recall of \modelRecall\%. This value is significantly greater than the one computed when testing the model in Section~\ref{sec:machine-learning}. Probably, this difference is explained by the fact that only projects that are completely unmaintained expose this situation in their READMEs. Therefore, it is easier to detect and identify this condition.

\begin{formal}
By validating the proposed model with projects that declare themselves as unmaintained, we achieve a {\bf recall} of \modelRecall\%.
\end{formal}

\vspace*{0.2cm}
\noindent{\bf \large RQ3: How early can we detect unmaintained projects?}\\[-.2cm]

\begin{figure}[!t]
\centering
\includegraphics[width=5cm]{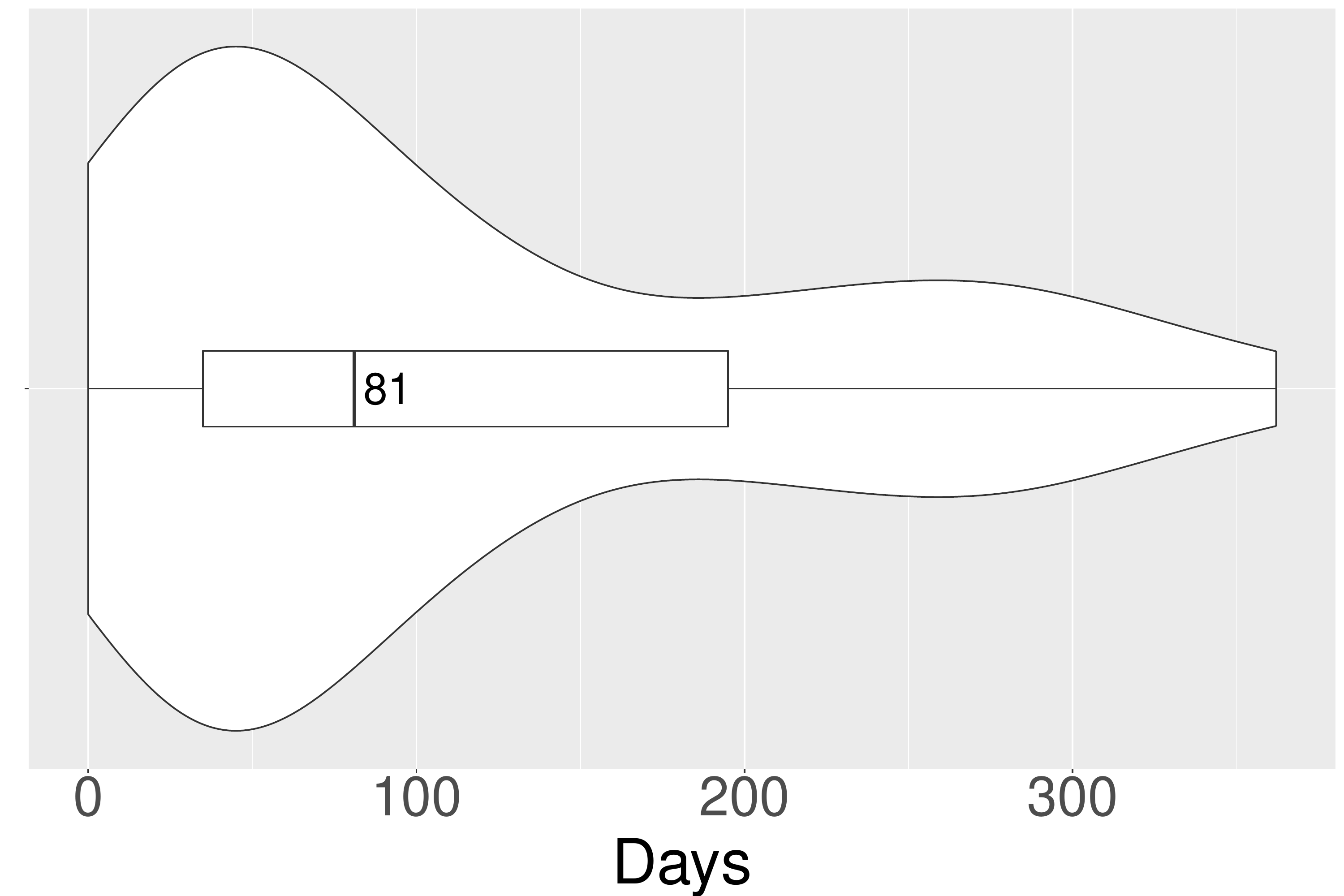}
\caption{Days since last commit for projects classified as unmaintained (considering the date of November, 2017, when the proposed model was computed).}
\label{fig:lma}
\end{figure}

\totalProjectsWithCommitsInTheLastYear\ (out of \totalTruePositivesAnswersBySurvey) projects classified as true positives by the surveyed developers have commits after November, 2016. Therefore, these projects would not be classified as unmaintained using the strategy followed in the literature, which requires one year of commit inactivity. In other words, in November, 2017, the proposed model classified \totalProjectsWithCommitsInTheLastYear\ projects as unmaintained, despite the existence of recent commits, with less than one year. Figure~\ref{fig:lma} shows a violin plot with the age of such commits, considering the date of November, 2017. The first, second, and third quartiles are 35, 81, and 195. Interestingly, for two projects the last commit occurred exactly on November, 30, 2018. Despite this fact, the proposed model classified these projects as unmaintained in the same date. If we relied on the standard threshold of one year without commits, these projects would have had to wait one year to receive this classification.

\begin{formal}
75\% of the studied projects are classified as unmaintained despite having recent commits, performed in the last year.
\end{formal}

\setcounter{figure}{5}
  
\begin{figure*}[!ht]
\centering

\subfloat[ref1][Commits]
{
\includegraphics[width=8cm]{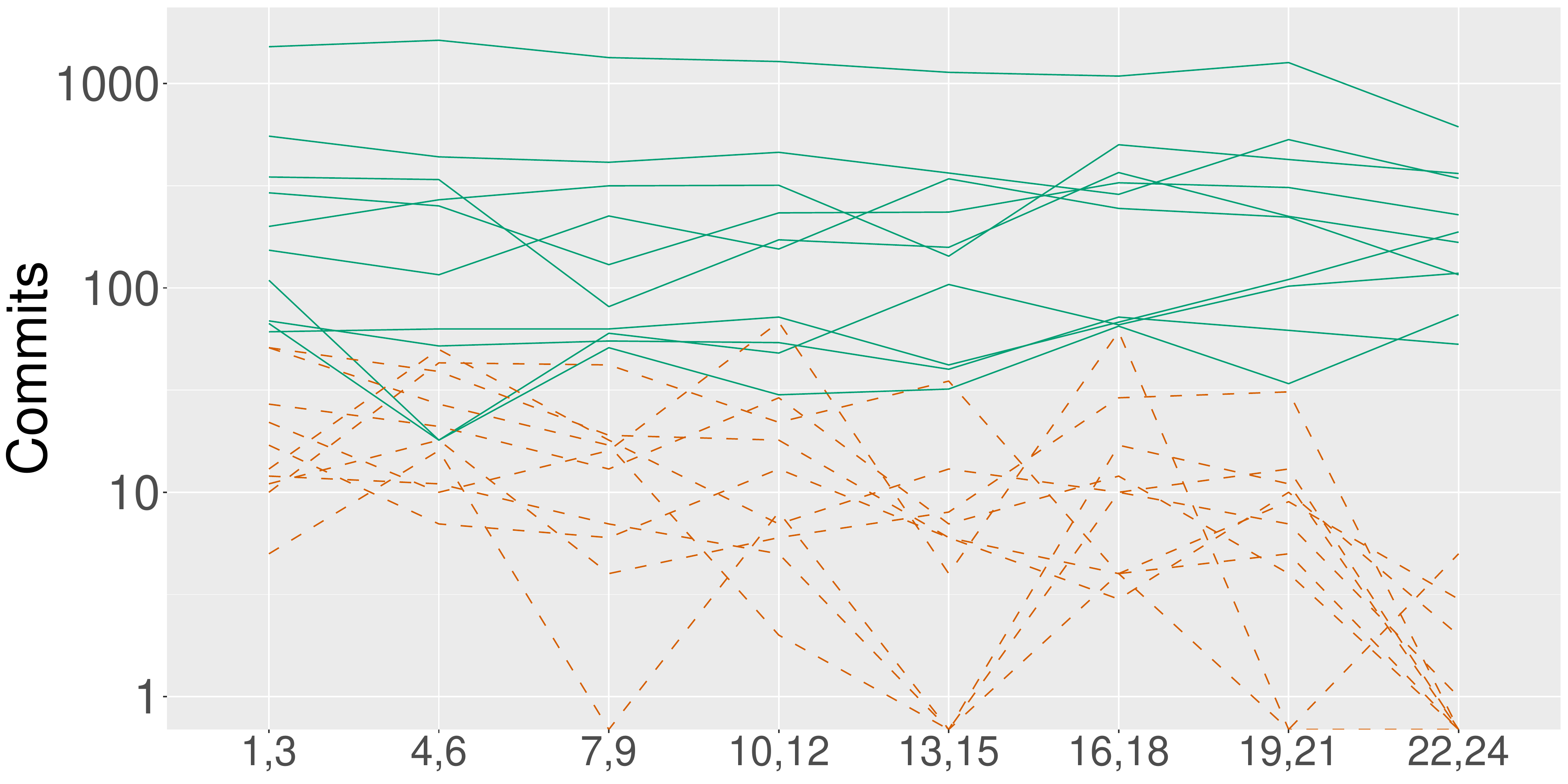}
}
\subfloat[ref2][Issues]
{
\includegraphics[width=8cm]{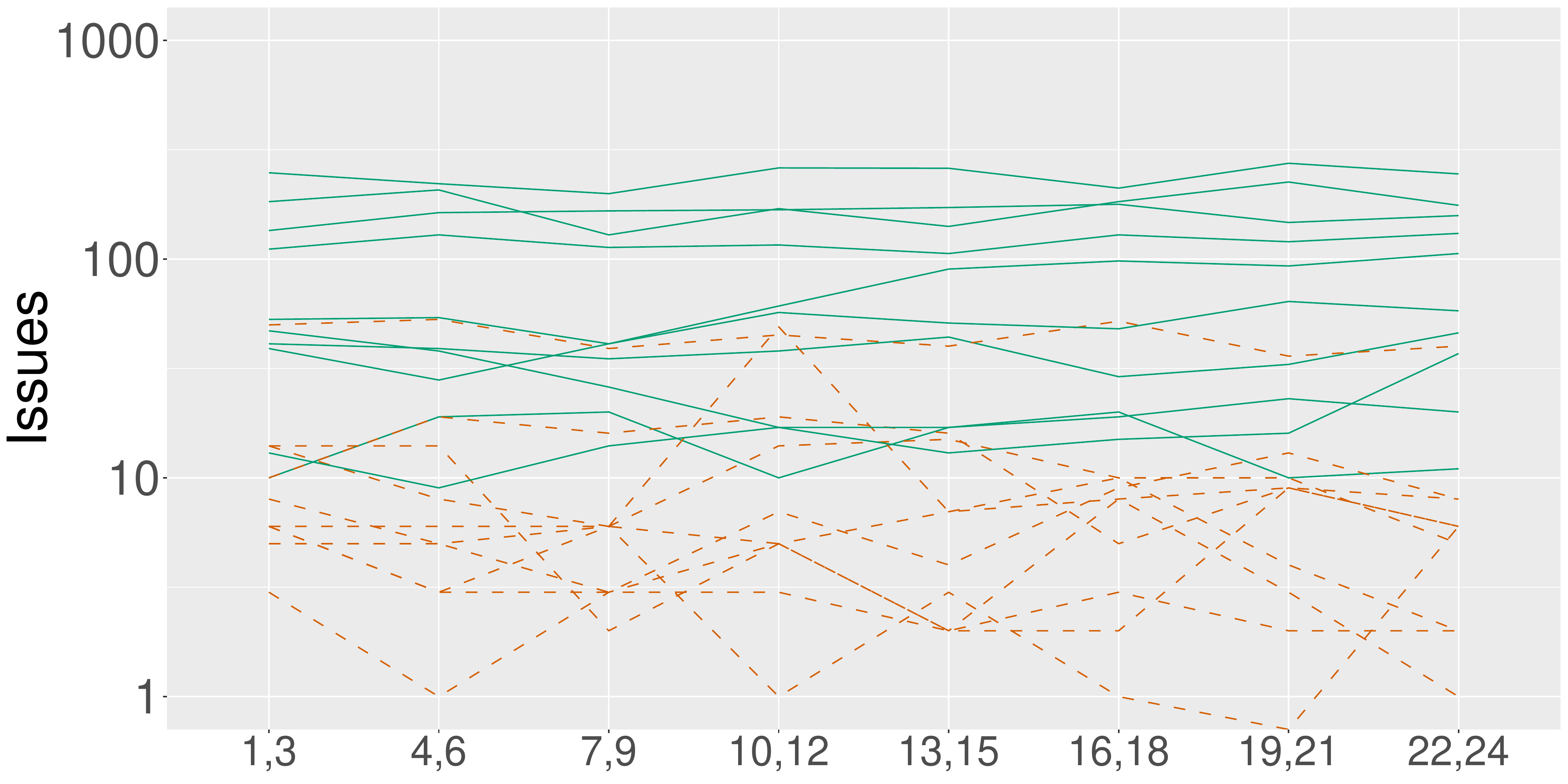}
}
\quad
\quad
\subfloat[ref3][Pull requests]
{
\includegraphics[width=8cm]{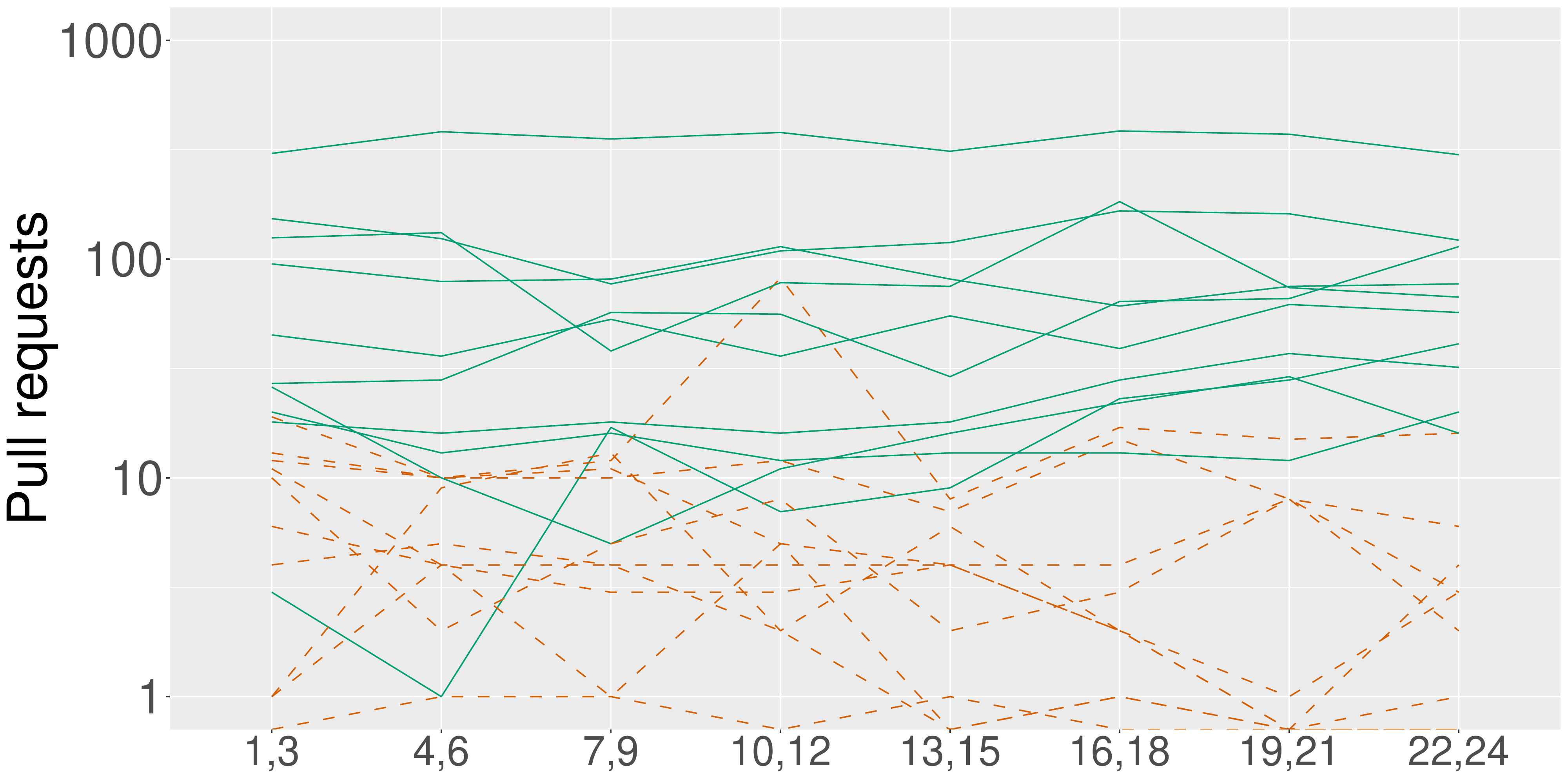}
}
\subfloat[ref4][Forks]
{
\includegraphics[width=8cm]{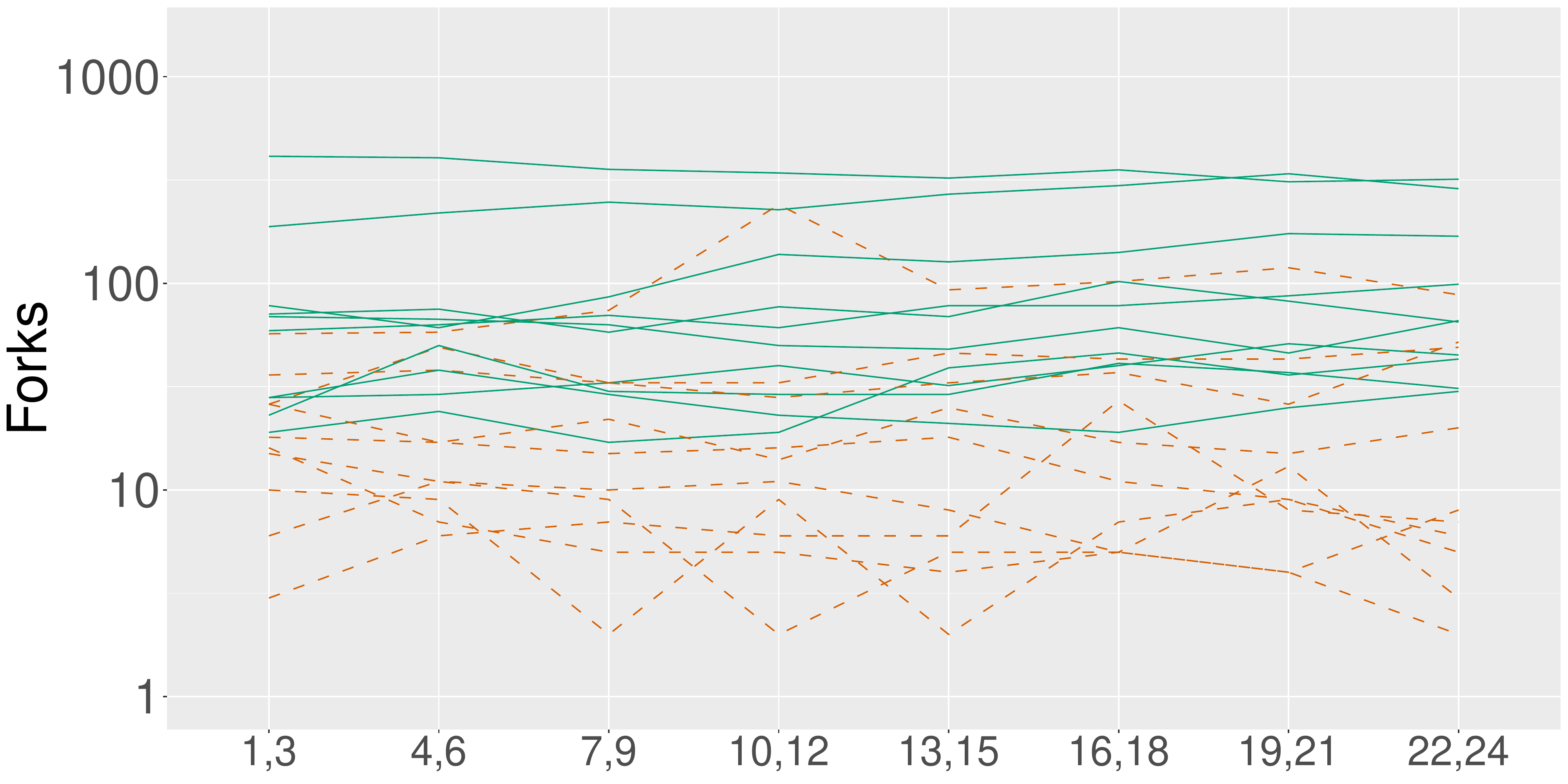}
}
\caption{Number of commits, issues, pull request, and forks over time of ten projects with maximal LMA (green lines) and ten projects with the lowest LMA in our dataset (red, dashed lines). Metrics are collected in intervals of 3 months (x-axis).}
\label{fig:active-vs-deprecated}
\end{figure*}

\setcounter{figure}{4}

\section{Level of Maintenance Activity}
\label{sec:level-of-maintenance-activity}

In this section, we define a metric to express the {\em level of maintenance activity} of GitHub projects, i.e., a metric that reveals how often a project is being maintained. The goal is to alert users about projects that although classified as under maintenance by the proposed model are indeed close to an unmaintained status.

\subsection{Definition}
\label{sec:walking-dead-methodology}

The proposed machine learning model---generated by Random Forest ---consists of multiple decision trees built randomly. Each tree in the ensemble determines a prediction to a target instance and the most voted class is considered as the final output. One possible prediction type of the Random Forest is the matrix of class probabilities. This matrix represents the proportion of the trees' votes. For example, projects predicted as {\em under maintenance} have probability $p$ ranging from 0.5 to 1.0. If  $p = 0.5$, the project is very similar to an unmaintained project; by contrast, $p = 1.0$ means the project is actively maintained. Using these probabilities, we define the {\em level of maintenance activity (LMA)} of a GitHub project as follows:
\[ LMA  =  2 * (p - 0.5) * 100 \]
This equation simply converts the probabilities $p$ computed by Random Forest to a range from 0 to 100; LMA equals to 0 means the project is very close to an unmaintained classification (since $p = 0.5$); and LMA equals to 100 denotes a project that is actively maintained (since $p = 1.0$). 

\subsection{Results}
\label{sec:walking-dead-results}
\begin{figure}[!ht]
\centering
\includegraphics[width=5cm]{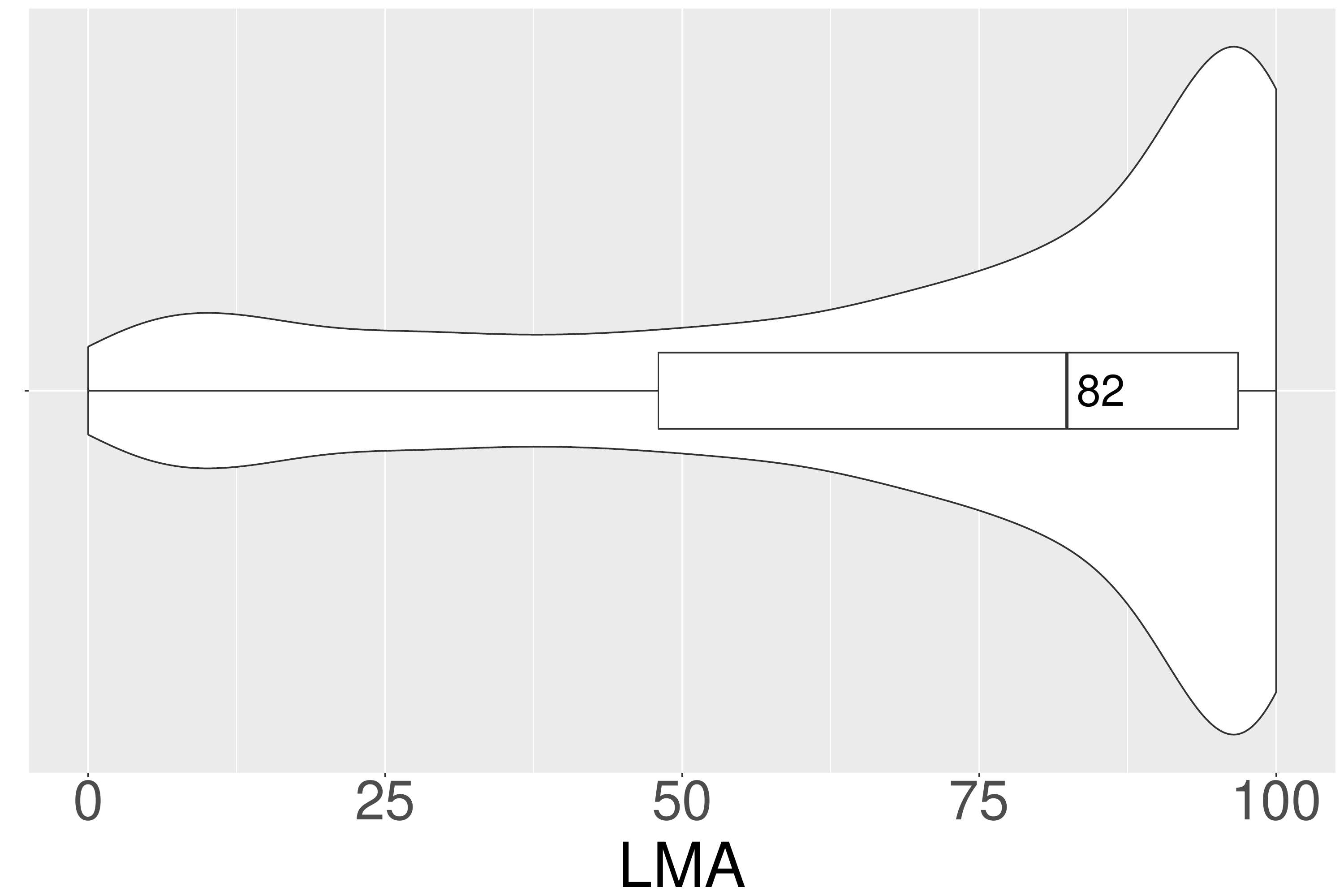}
\caption{Level of maintenance activity (LMA).}
\label{fig:active_projects_score}
%\vspace{-0.4cm}
\end{figure}

Figure~\ref{fig:active_projects_score} shows the LMA values for each project predicted as {\em under maintenance} (\testsetClassifiedAsActive\ projects, after excluding the projects used to train and test the proposed model, in Section~\ref{sec:machine-learning}). The first, second, and third quartiles are 48, 82, and 97, respectively. In other words, most studied projects are under constant maintenance (median 82). Indeed, 171  projects (5.8\%) have a maximal LMA, equal to 100. This list includes well-known and popular projects such as {\sc twbs/bootstrap},  {\sc meteor/meteor}, {\sc rails/rails}, {\sc webpack/web\-pack}, and {\sc elastic/e\-las\-ticsearch}.

\setcounter{figure}{6}

\begin{figure*}[!ht]
\centering

\subfloat[ref1][LMA vs Stars ($\rho$ = 0.10)]
{
\includegraphics[width=8cm]{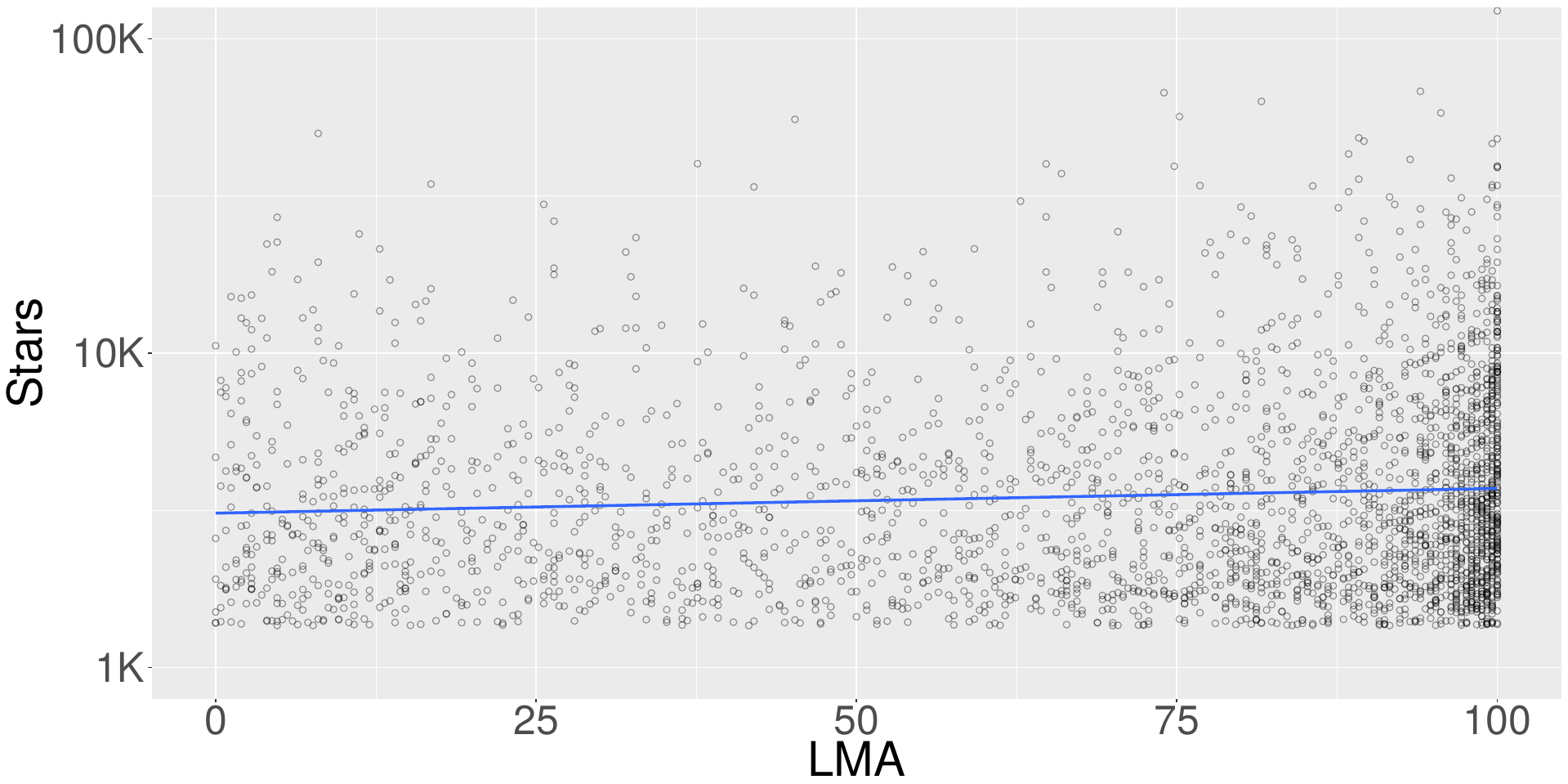}
}
\subfloat[ref2][LMA vs Contributors ($\rho$ = 0.44)]
{
\includegraphics[width=8cm]{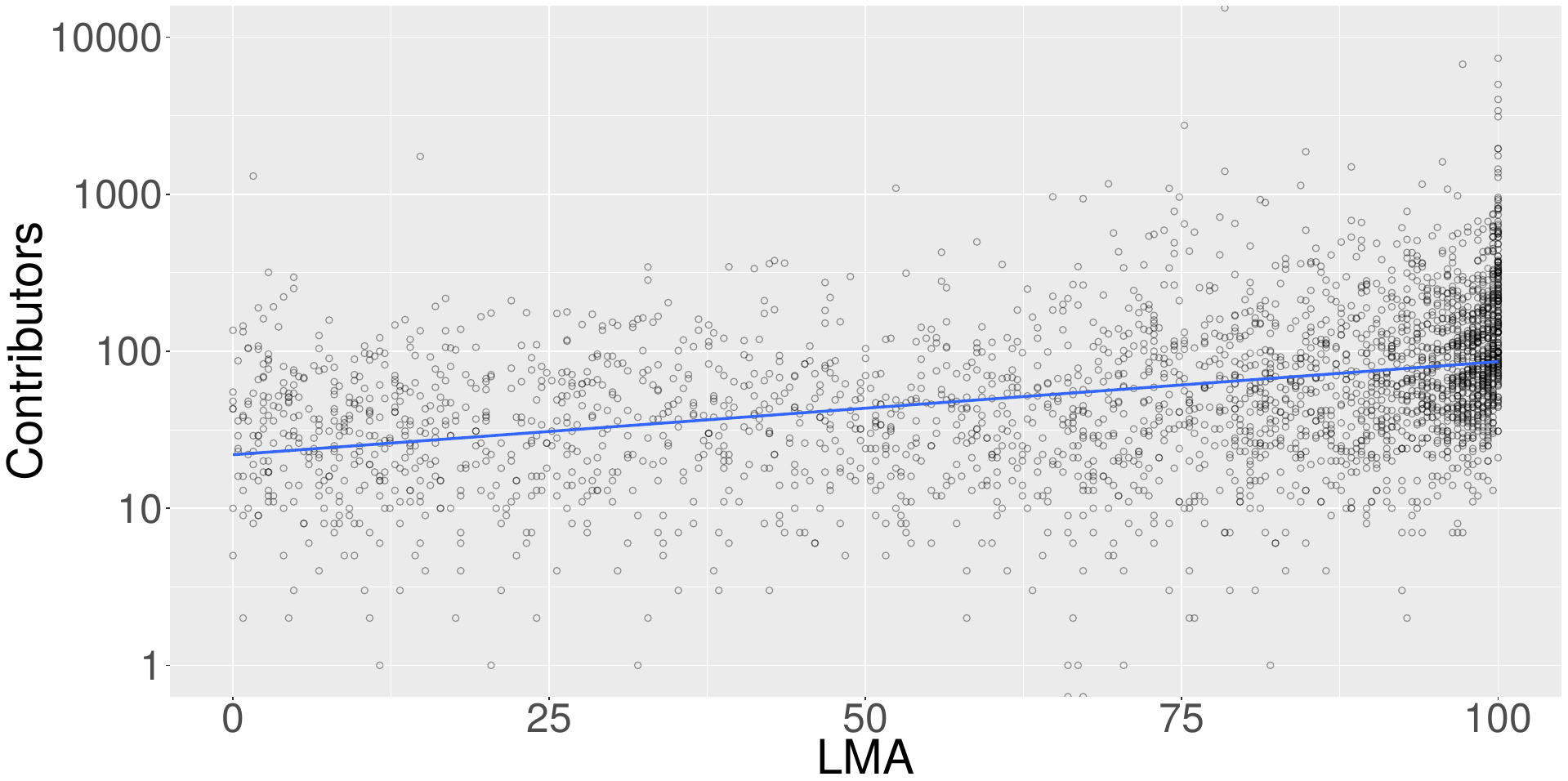}
}
\quad
\quad
\subfloat[ref3][LMA vs Core contributors ($\rho$ = 0.15)]
{
\includegraphics[width=8cm]{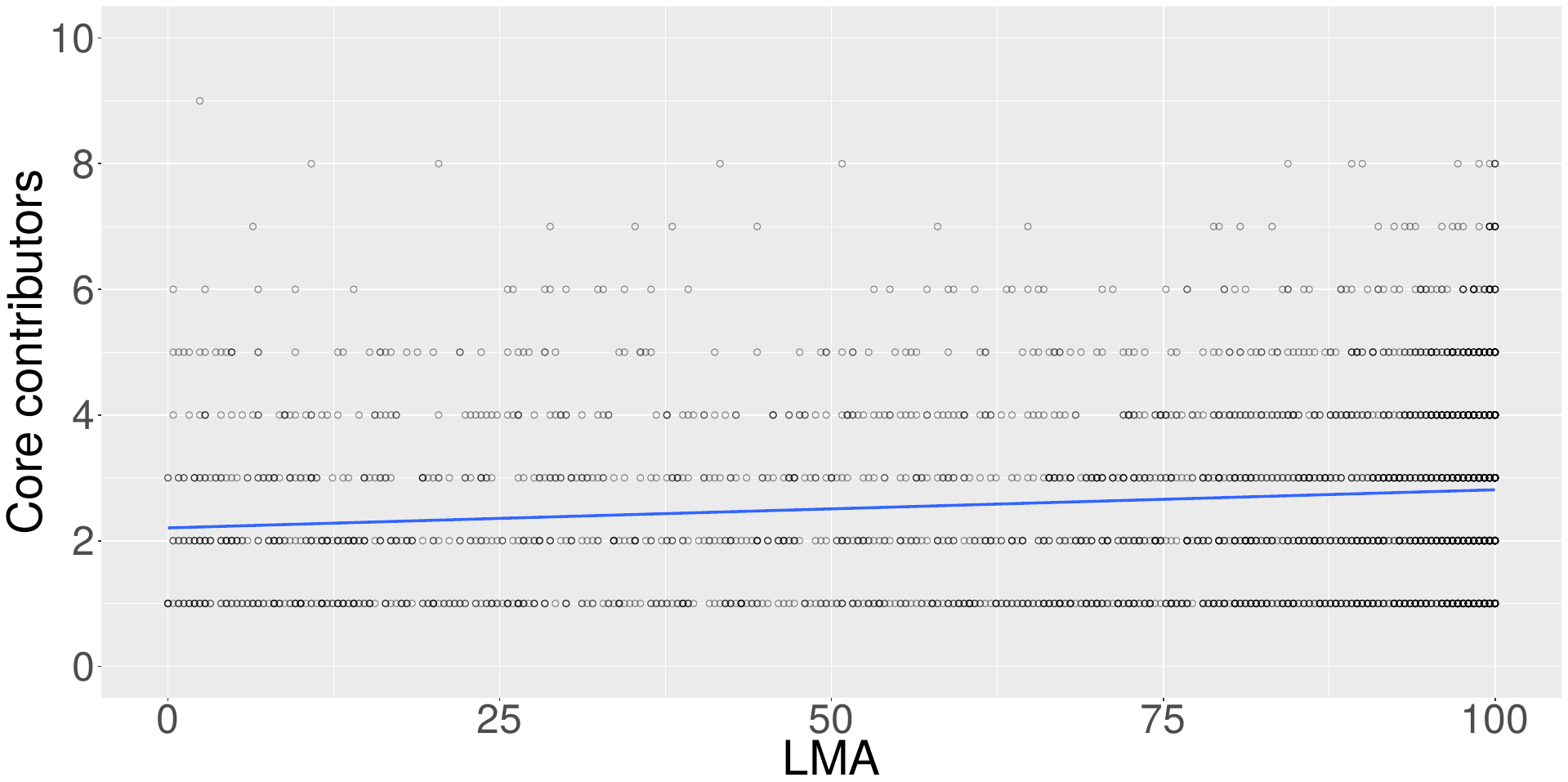}
}
\subfloat[ref4][LMA vs LOC ($\rho$ = 0.38)]
{
\includegraphics[width=8cm]{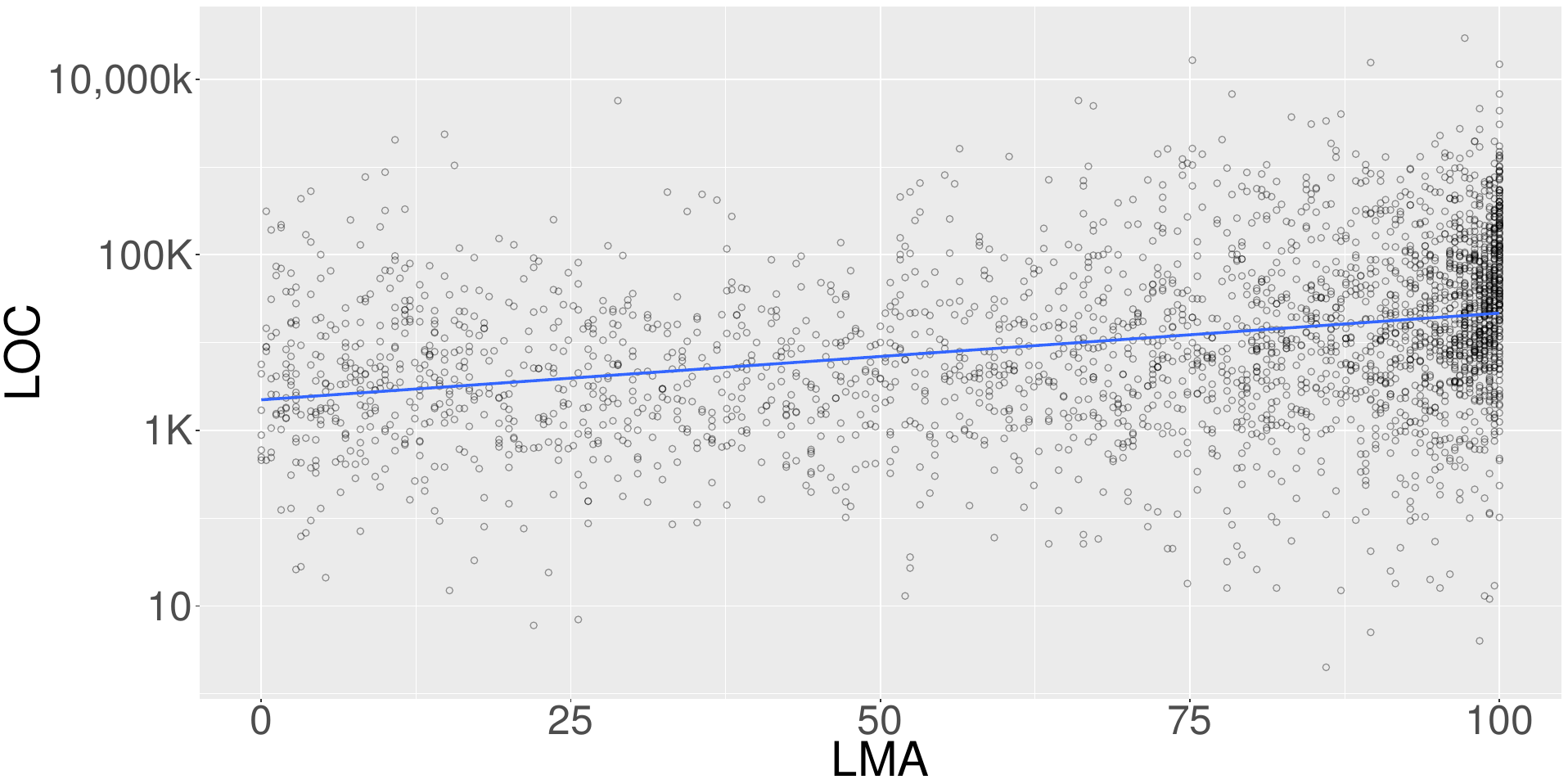}
}
\caption{Correlating LMA with (a) stars, (b) contributors, (c) core contributors, and (d) size. Spearman's $\rho$ is also presented.}
\label{fig:lma-vs-features}
\end{figure*}

Figure~\ref{fig:active-vs-deprecated} compares a random sample of  10 projects with LMA equals to 100 (actively maintained, therefore) with ten projects with the lowest LMA (0 $\leq$ LMA $\leq$ 0.4). These projects are compared using  number of commits (Figure~\ref{fig:active-vs-deprecated}a), number of issues (Figure~\ref{fig:active-vs-deprecated}b), number of pull requests (Figure~\ref{fig:active-vs-deprecated}c), and number of forks (Figure~\ref{fig:active-vs-deprecated}d), in the last 24 months. Each line represents the project's metric values. The figures reveal major differences among the projects, regarding these metrics. Usually, the projects with high LMA present high values for the four  considered metrics (commits, issues, pull requests, and forks), when compared with projects with low LMA. In other words, the figures suggest that LMA plays an aggregator role of maintenance activity over time.

Figure~\ref{fig:lma-vs-features} shows scatterplots correlating LMA and number of stars, contributors, core contributors, and size (in LOC) of projects classified as under maintenance. To identify core contributors, we use the most common heuristic described in the literature: core contributors are the ones responsible together for at least 80\% of the commits in a project~\cite{koch2002effort,mockus2002two,robles2009evolution}. To measure the size of the projects, in lines of code, we used the tool {\sc AlDanial/cloc}\footnote{https://github.com/AlDanial/cloc}, considering only the programming languages in the TIOBE list.\footnote{https://www.tiobe.com/tiobe-index} We also compute Spearman’s rank correlation test for each figure. The correlation with stars and with core contributors is very weak ($\rho$ = 0.10 and $\rho$ = 0.15, respectively); with size, the correlation is weak ($\rho$ = 0.38); and with contributors, it is moderate ($\rho$ = 0.44); all p-values are less than 0.01. Therefore, it is common to have highly popular projects, by number of stars, presenting both low and high LMA values. For example, one project has 50,034 stars, but LMA = 8. A similar effect happens with size. For example, one project has $\approx$2 MLOC, but LMA = 10.8. The highest correlation is observed with contributors, i.e.,~projects with more contributors tend to have higher levels of maintenance activity.

\subsection{Validation with False Negative Projects}

In Section~\ref{sec:validation}, we found four projects that although declared by their developers as {\em unmaintained} are predicted by the proposed machine learning model as {\em under maintenance}. Therefore, these projects are considered false negatives, when computing recall. Two of such projects has a very low LMA:
{\sc nicklockwood/iRate} (LMA = 2) and {\sc gorangajic/react-icons}	(LMA = 12). Therefore, although predicted as {\em under maintenance},  this project is similar to projects classified as {\em unmaintained}, as suggested by its low LMA.
A second project has an intermediate LMA value: {\sc spotify/HubFramework}	(LMA = 39.2). Finally, one project {\sc Homebrew/homebrew-php} has a high LMA value (LMA = 99.2). However, this project was migrated to another repository, when facing continuous maintenance.
In other words, in this case, the GitHub repository was deprecated, but not the project; therefore, {\sc Homebrew/homebrew-php} is a false, false negative (or a true negative).

\section{Threats To Validity}
\label{sec:threats}

The threats to validity of this work are described as follows:\\[-.25cm]

\noindent{\bf External Validity:} Our work examines open source projects on GitHub. We recognize that there are popular projects in other platforms (e.g., Bitbucket, SourceForge, and GitLab) or projects that have their own version control installations. Thus, our findings may not generalize to other open source or commercial systems. A second threat relates to the features we have considered. By adding other features, we may improve  the prediction of unmaintained projects; however, given our high prediction performance, we feel confident that our features are effective. Also, some of the features we use may not be available in other projects, however, most of our features are available in most code control repositories/ecosystems. In the future, we intend to investigate additional projects and consider more features.\\[-.25cm]

\noindent{\bf Internal Validity:} The first threat relates to the selection of the survey participants. We surveyed the project owner, in the case of repositories owned by individuals, or the developer with the highest number of commits, in the case of repositories owned by organizations. We believe the developers who replied to our survey are the most relevant given their level of activity in the project. It is also possible that most missing answers are from developers of unmaintained projects. As a second threat, the themes of the survey were defined and organized by the authors of the paper. As with any human activity, the derived themes may be subject to bias and different researchers might reach different observations. However, to mitigate this threat, a first choice of themes was conducted in parallel by the first two authors of this paper. Also, they attended several meetings during a whole week to improve the initial selected themes. A third threat relates to the parameters used to perform our experiment. We set the number of trees to 100 to train our classifier. To attenuate the bias of our results, we run 5-fold cross-validation and use the average performance for 100 rounds. A forth threat is related to how the accuracy of our machine learning approach was evaluated. We relied on developer replies about their projects to evaluate the performance of our machine learning classifier. In some cases, the developer replies (or developers who did not reply) may impact our results. That said, our survey had a response rate of 37.1\%, which is very high for a software engineering study, giving us confidence in the reported performance results.\\[-.25cm]

\noindent{\bf Construct Validity:} A first threat relates to the definition of active projects. We consider as active projects those with at least one release in the last month (Section~\ref{sec:machine-learning}). We acknowledge a threat in the definition of the time frame. To mitigate this threat, the first paper's author inspected each selected project to look for deprecated projects (\totalProjectsWithDeprecatedMessageInReadmeOfSurvey\ projects declare they are no longer being maintained) and we conduct a survey with \totalSurveyAnswersByDevelopers\ developers to confirm our findings. A second threat is related to the projects we studied. Our dataset is composed of the most starred projects (and additional filtering). Although the starred projects may not be representative of all open source projects, we did carefully select such projects to ensure that our study is conducted on real (and not toy) projects.  

\section{Related Work}
\label{sec:related-work}

\noindent{\bf Machine Learning.} 
Recently, the application of machine learning in software engineering contexts has gained much attention. Several researchers have used machine learning to accurately predict defects (e.g.~\cite{peters2013better}), improve issue integration (e.g.,~\cite{Alencar2014}), enhance software maintenance (e.g.,~\cite{gousios2014exploratory}), and examine developer turnover (e.g.,~\cite{bao2017will}). For example, \citet{gousios2014exploratory} investigate the use of machine learning to predict whether a pull request will be merged. They extract 12 features organized into three dimensions: pull request, project, and developer. They conduct their study using six algorithms (Logistic Regression, Naive Bayes, Decision Trees, AdaBoost with Decision Trees, and Random Forest). \citet{bao2017will} build a model to predict developer turnover, i.e., whether a developer will leave the company after a period of time. They collect several features based on developers monthly report from two companies. The authors evaluate the performance of five classifiers (KNN, Naive Bayes, SVM, Decision Trees, and Random Forest). In both studies, Random Forest outperforms the results of other algorithms. In another study, \citet{martin2016causal} train a Bayesian model  to support app developers on causal impact analysis of releases. They mine time-series data about Google Play app over a period of 12 months and survey developers of significant releases to check their results.
\citet{tian2015characteristics} use Random Forest to predict whether an app will be high-rated. They extract 28 factors from eight dimensions, such as app size and library quality. Their findings show that external factors (e.g., number of promotional images) are the most influential factors. Our study also uses machine learning techniques, however, our main goal is to detect projects that are not going to be actively maintained. Moreover, our study extracts project, contributor and owner features that we input to the machine learning models.\\[-.25cm]

\noindent{\bf Open source projects maintainability.} In previous work~\citep{coelho2017why}, we survey maintainers of 104 failed open source projects to understand the rationale for such failures. Their findings revealed that projects fail due to reasons associated with project properties (e.g., low maintainability), project team (e.g., lack of time of the main contributor), and to environment reasons (e.g., project was usurped by a competitor). Later, we report results of a survey with 52 developers who recently became core contributors on popular GitHub projects~\citep{coelho2018why}. Our results show the developer's motivations to assume an important role in FLOSS projects (e.g., to improve the projects because they use them), the project characteristics (e.g., a friendly community), and the obstacles they faced (e.g., lack of time of the project leaders). 

Also related is the work by \citet{yamashita2014magnet}, which adapts two population migration metrics in the context of open source projects. Their analysis enables the detection of projects that may become obsolete.  \citet{khondhu2013all} report that more than 10,000 projects are inactive on SourceForge. They use the maintainability index (MI)~\cite{oman1992metrics} to compare the maintainability between inactive projects and projects with different statuses (active and dormant). Their results reveal that the majority of inactive systems are abandoned with a similar or increased maintainability, when compared to their initial status. Nonetheless, there are critical concerns on using MI as a predictor of maintainability~\cite{bijlsma2012faster}. \citet{nadia2016roads} reports risks and challenges to maintain modern open source projects. She argues that open source plays a key role in the digital infrastructure of our society today. Opposed to physical infrastructure (e.g., bridges and roads), open source projects still lack a reliable and sustainable source of funding. Other recent research on open source has focused on the organization of successful open source projects~\cite{mockus2002two} and on how to attract and retain contributors~\cite{zhou2015will, steinmacher2016overcoming, leeunderstanding, pinto2016more, canfora2012going}. Our work enhances the aforementioned work by contributing factors and proposing the use of machine learning to accurately identify projects that are not going to be actively maintained.  

\section{Conclusion}
\label{sec:conclusion}

In this paper, we proposed a machine learning model to identify  unmaintained GitHub projects. By our definition, this status includes three types of projects: finished projects, deprecated projects, and stalled projects. We validated the proposed model with the principal developers of \totalSurveyAnswersAndReadme\ projects, achieving a precision of \modelPrecision\% (RQ1). Then, we  used the model with \totalProjectsUnmaintainedByReadme\ deprecated projects---as explicitly mentioned in their GitHub page. In this case, we achieved a recall of \modelRecall\% (RQ3). We also showed that the proposed model can identify unmaintained projects early, without having to wait for one year of inactivity, as commonly proposed in the literature (RQ3). Finally, we defined a metric, called Level of Maintenance Activity (LMA), to assess the risks of projects become unmaintained. We provided evidence on the applicability of this metric, by investigating its usage in \testsetClassifiedAsActive\ projects classified as under maintenance, in our dataset.

Due to its high accuracy (precision= \modelPrecision\% and recall= \modelRecall\%), the model proposed in this paper can be used by developers to check the maintenance status of an open source project, before deciding to use it. This information has a key value, since there is a growing concern on the sustainability of modern open source projects~\cite{nadia2016roads}.

As future work, we intend to implement a tool to provide information about the maintenance status and the level of maintenance activity of open source projects. The dataset used in this paper is available at: \url{https://zenodo.org/record/1313637}. Due to privacy concerns, we are omitting the name of the unmaintained projects.

\section*{Acknowledgments}
Our research is supported by CAPES, FAPEMIG, and CNPq. We would also like to thank the \totalSurveyAnswersByDevelopers\ GitHub developers who kindly answered our survey.

\balance
\bibliographystyle{ACM-Reference-Format}
\bibliography{esem2018}

\end{document}